# Nanocrystal tuned ammonia gas sensing technique via impedance spectroscopy

Neha Sharma[1], Debanjan Bhattacharjee[2], Sunita Kumari[3], Sandip Paul Choudhury[1,*]

[1]*Amity School of Applied Sciences, Amity University Rajasthan, Jaipur, Rajasthan – 303002, India*

[2] *Department of Physics, Manipal University Jaipur, Rajasthan - 303007, India*

[3] *Department of Physics, Indian Institute of Technology, Jodhpur, Rajasthan - 342037, India*

[*]*Corresponding author email: sandip.pchoudhury@gmail.com*

**Abstract:**

Ammonia is a harmful chemical hazard known for its widespread industrial use. Exposure to ammonia can cause environmental damage, human health hazards, and huge economic losses. Therefore, ammonia gas sensors are essential for detecting ammonia leaks to avoid serious accidental injury and death. In this study, we synthesize a nanostructured ($WO_3$) n-type metal oxide semiconductor doped with a rare earth element-transition metal (Ce-Cu) via hydrothermal method for ammonia gas sensing application. Structural analysis was performed using XRD and FESEM. Further we investigate the optical properties via UV-visible spectroscopy, FTIR, and PL. We found that doping of (Ce-Cu) led to significant improvement in thermal stability for ammonia detection and selectivity performance compared to that pure one, across a wide frequency range. We believe that these studies will pave the way for exploring the use of Ce-Cu to improve the gas sensing properties of semiconductor-based gas sensors.

## 1. Introduction:

Ammonia and volatile organic compound (VOC) gases cause environmental pollution and adversely affect human health [1]. Ammonia and VOCs, such as methanol, formaldehyde and toluene, are colourless and volatile liquids. Exposure to such gases causes several health problems including headache, vomiting and breathing problems. Excessive exposure to these gases can also cause leukaemia and cancer [2]. Notably, ammonia is one of the species present in the environment that originates from the chemical, pesticide and fertilizer industries. Researchers have developed several methods for gas removal including chromatography, colorimetric methods, spectrophotometry and chemical sensors for detecting gases [3]. Chemical sensors stand out among the methods due to their affordability, portability and ease of use.

Metal oxide semiconductors (MOSs) are a type of chemical sensors. Metal oxide semiconductors (MOSs) have great potential in the field of material science. There are many MOSs such as ZnO, SnO$_2$, TiO$_2$, and WO$_3$. WO$_3$ is a n-type metal oxide semiconductor with a bandgap of 2.6-3.0 eV. Tungsten oxide have cubic, triclinic, monoclinic, orthorhombic, tetragonal, and hexagonal structures at different temperature [4]. Because of its excellent surface permeability and high specific surface area, nanostructured WO$_3$ is apt for multiple applications. Many morphologies of WO$_3$ nanostructures, such as nanorods (NRs), nanosheets, and thin films (TFs), have been synthesized for a range of application, which includes memory devices [5], photodetectors [6], gas sensors [7], effective water splitting [8, 9], photo electrocatalytic activity [9] and high-temperature diodes[10]. WO$_3$ sub-stoichiometric oxide (WO$_{3-x}$) is frequently observed in WO$_3$ because of many oxygen deficits, including WO$_{2.9}$, WO$_{2.83}$, WO$_{2.8}$, and WO$_{2.72}$. In other words, the WO$_{3-x}$ lattice may support a significant amount of oxygen vacancy and have a lot of W$^{5+}$. As a result, altering the oxygen vacancies in WO$_{3-x}$ may be able to significantly alter the electron density and conductivity. So, the small change in oxygen vacancy of WO$_3$ can also change the conductivity of the tungsten oxide.

Tungsten oxide is widely used as electrochromic, photochromic and gas-sensing materials. Efkere et al [11] investigated how WO$_3$ nanostructures deposited on Si-substrate by RF magnetron sputtering at different annealed temperatures modify their structural, optical and morphological properties. They also reported that as the annealing temperature increases from 300-700°C, the value of band gap decreases from 3.44-3.15 eV. Doping is an efficient way to enhance the electrical and optical properties of tungsten oxide. When metal oxide semiconductors such as WO$_3$, ZnO, SnO2 and CuO are doped with transition metals and rare earth elements, there electronic and optical properties changes. These changes are caused by the creation of new defect states in the band structure, modifying the optical band gap of the doped material. Rajendran et. al [12] have synthesized the pure and Cu-doped WO$_3$ by the co-precipitation method and investigated that after doping of Cu in WO$_3$ the morphology changes from nanoplates to nanorods. Sriram et. al [13] synthesized pure WO$_3$, 2% Cu-doped WO$_3$, 4% Cu-doped WO$_3$, and 6% Cu-doped WO$_3$ thin films. Based on structural, morphological, and chemical characterization, they observed that among these samples, 4% Cu-doped WO$_3$ exhibited an increase in oxygen vacancies and surface area. They also reported that the response values of pure $WO_3$, 2% Cu-doped $WO_3$, 4% Cu-doped $WO_3$, and 6% Cu-doped $WO_3$ towards 50 ppm ammonia gas at room temperature were 4.2, 4.6, 8.38, and 7, respectively.

The rare earth elements (Eu, Tb, and Ce) in $WO_3$ are an effective method to enhance electrical and photoluminescence properties at room temperature. Wang et. al [14] have synthesized REE (Eu, Sm, Ce, and Gd) doped $WO_3$ by one-step hydrothermal method and found that rare earth elements doping modified their physical and morphological phases. They also reported that after REEs-doped $WO_3$ nanofilm has greater optical contrast and high electrochromic properties. Govindaraj et. al [15] reported that doping of Gd in $WO_3$ decreases the crystalline size and reduces the band gap from 2.80 to 2.64 eV. They also reported that 5% Gd-doped $WO_3$ NR (nanorod) has better photocatalytic dye degradation activity. Mohanraj et. al [16] reported that Ce dopant 0 to 5% in $WO_3$ decreases the crystallite size from 37 to 31nm and also reduces the band gap value. They also reported that 5%Ce-doped $WO_3$ nanoparticles have a maximum conductivity of $5.01 \times 10^{-7}$ S/cm and show highest photocatalytic activity. Liu et al. [17] synthesized mesoporous Ce-doped $WO_3$ using a facile in situ cooperative assembly method combined with a carbon-supported crystallization strategy for detecting trace gases at low temperatures. Ce doping in mesoporous $WO_3$ increases oxygen vacancies on the $WO_3$ lattice surface. It also enhances the specific surface area, ranging from 59 to 72 m²/g. They reported that the mesoporous walls of Ce-doped $WO_3$ exhibit a high response value of 381 for 50 ppm $H_2S$ gas at an operating temperature of 150°C. DFT calculations further revealed that Ce-doped $WO_3$ has higher adsorption energy and greater electron transfer compared to pure $WO_3$ for $H_2S$ gas adsorption.

The ternary compounds are a new family of material being explored by present researchers due to their higher surface area and higher catalytic properties. Tran et. al [18] reported that Ag@rGO@$WO_3$ enhances the photocatalytic property of pure $WO_3$ and also reduces the optical band gap. Mohammadi et. al [19] have synthesized the pure, Zn, Cu and Zn-Cu doped $WO_3$ samples respectively by the precipitation and co-precipitation method and found that the doping affects the optical properties. They also found optical band gap values for these four samples are 3.2, 3.12, 3.08, and 2.97 eV respectively, and photocatalytic property increases for the Zn-Cu co-dopped sample. These four samples have good antimicrobial properties.

There are various methods to synthesize pure and doped $WO_3$ such as chemical co-precipitation method [20], one-step hydrothermal method [21], sol-gel method [22], solvothermal method [23] and wet-chemical method [24]. From these methods, the hydrothermal method is a cost-effective method to synthesize pure, Cu-doped, Ce-doped and Cu-Ce-doped $WO_3$, respectively. Impedance plots can be used to analyse the AC electrical behaviour of pure and doped tungsten oxide nanocrystals and the gas sensing behaviour of pure $WO_3$ and doped-$WO_3$. Impedance plots are plots between the imaginary part of the impedance ($Z''$), which represents the reactance, and the real part of the impedance ($Z'$), which represents the resistance. This technique is helpful in understanding the behaviour of individual properties of sensing materials, including grain boundaries, grain size, and the boundary between the electrode and the sensing material.

There are many reports on the optical, electronic and structural properties of Ce and Cu-doped $WO_3$. However, to the best of our knowledge, no research has been done on Ce-Cu doped $WO_3$ for structural, optical and electronic properties. Researchers have used the ternary compound to enhance the properties such as electrical, optical, magnetic and photocatalytic properties of MOS. In this work, we study the effect of Ce-Cu-doped $WO_3$ for gas sensing application. The gas sensing performance is investigated via impedance spectroscopy, focusing on the sensitivity and selectivity of ammonia at different temperatures.

## 2. Experimental details

### 2.1 Chemicals

The chemicals used for the synthesis of the samples are tungsten hexachloride ($WCl_6$), cerium nitrate hexahydrate (Ce $(NO_3)_3.6H_2O$), copper chloride ($CuCl_2$) and ethylene glycol. All chemical is procured from Sigma Aldrich and are of pure analytical grade. Pure $WO_3$, Ce doped $WO_3$, Cu-doped $WO_3$, and ternary compound Ce-Cu-doped $WO_3$ nanoparticles were synthesized by the hydrothermal method. The Tungsten hexachloride ($WCl_6$), Cerium nitrate hexahydrate (Ce $(NO_3)_3.6H_2O$), and Copper chloride ($CuCl_2$) were used for tungsten, Cerium, and Copper sources, respectively. Pure $WO_3$, Ce and Cu-doped $WO_3$ and Ce-Cu-doped $WO_3$ are coded as S1, S2, S3 and S4, respectively.

### 2.2 Synthesis method

For the preparation of S1, the tungsten hexachloride ($WCl_6$) was used as the tungsten source, and Ethylene glycol water as the solvent. Firstly, 1.09 gm $WCl_6$ was slowly added into 60 ml of ethylene glycol and 10 ml of water (EG-water) solvent in a beaker. Then to obtain the homogeneous solution the mixture was stirred at room temperature using a magnetic stirrer. After getting the homogeneous solution, the solution mixture was transferred into the 100 ml

Teflon-lined autoclave and kept at 200°C temperature for 5h in an electric oven. The autoclave was then cooled down to room temperature before obtaining the sample. The as-prepared sample was washed several times with ethanol and deionized water, dried the sample at 90°C temperature in an oven for 10h, and crushed in an agate mortar. Similar procedures were used to prepare S2, S3, and S4, except for adding the appropriate amounts of Ce $(NO_3)_2$.$6H_2O$ (2 wt%), $CuCl_2$ (1 wt%), and Ce $(NO_3)_2$.$6H_2O$ (2 wt%) $CuCl_2$ (1 wt%) to the solution mixture, respectively. The finely powdered samples (S1, S2, S3, S4) were annealed at 400°C temperature for 2h using a muffle furnace. The schematic diagram of synthesis process is shown in Figure 1.

## 2.3 Characterization:

The synthesized nanocrystal structures of the samples were analysed using various characterization tools to understand their behaviour. First, an X-ray diffractometer (XRD, Rigaku Corp., D/Tex Ultra 250) with CuKα radiation (40 kV and 40 mA) and a scanning rate of 10.00 °/min was employed to investigate the crystal phases and composition of the materials. The surface morphology of the synthesized samples was examined using a Field Emission Scanning Electron Microscope (FE-SEM, JEOL; JSM 7610FPLUS), and their elemental mapping was analysed through energy-dispersive spectroscopy (EDS, EDAX APEX). The band gap and absorption spectra of the synthesized samples were calculated using an ultraviolet-visible spectroscope. Fourier-transform infrared (FTIR) spectroscopy was used to identify the functional groups present in the samples. Photoluminescence (PL) spectroscopy was conducted to analyse the optical and electronic properties of the samples.

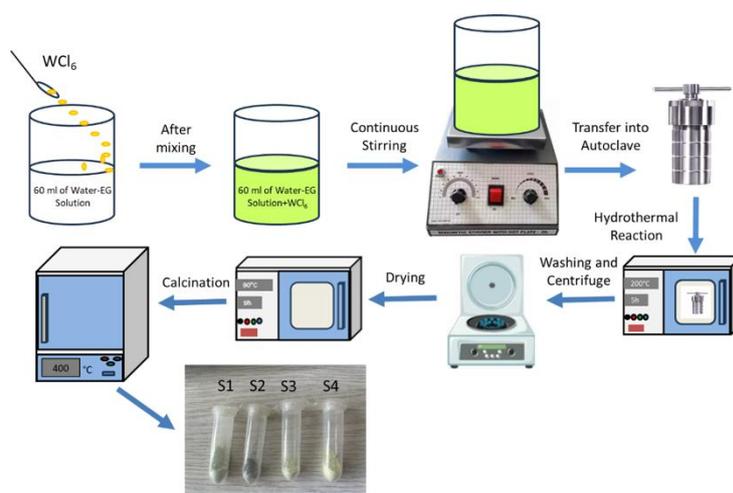

**Figure 1** Hydrothermal synthesis process for preparation of S1, S2, S3 and S4

## 3. Results and Discussion

### 3.1 Structural characterization

#### 3.1.1 XRD (X-ray diffraction)

For the structural analysis XRD were performed for S1, S2, S3 and S4, respectively and the XRD pattern of these are shown in Fig.2. The indexed (hkl) planes are (002), (020), (200), (120), (112), (022), (-202) and (400) are clearly observed that confirmed the monoclinic structure of the $WO_3$ nanocrystals with space group P2$_1$/n in JCPDS card number 01-072-0677. The sharp peaks in XRD pattern indicates that the formation of S1, S2, S3 and S4, respectively. For the Ce-doped $WO_3$ (S2), Cu-doped $WO_3$ (S3) and Ce-Cu-doped $WO_3$ (S4) the peak intensity increases in comparison to pure $WO_3$ (S1) that indicates that doping have an influence in the crystal growth [12]. The peak intensity increases due to the lattice distortion. This distortion is a result of difference in the ionic radius of host $W^{6+}$ (62 pm) and dopant ion ($Cu^{2+}$:73 pm; $Ce^{4+}$: 87 pm). Also, the peaks of S2 and S3 are more sharp than S1 that indicates the higher degree of crystallinity [25]. The peaks reveal broadness and lower intensity after incorporation of Ce-Cu both in $WO_3$. This indicates the decrease in the crystallite size of nanocrystals and due to the presence of micro strains. Due to mismatch between valency and atomic size of dopants in the host $WO_3$ material the micro strains increase which reduce the particle size.

There is a change in d-spacing value and lattice parameter after doping which indicates the presence of dopant atoms in $WO_3$ nanocrystals. The interplanar distance was calculated by Bragg's diffraction equation which is shown in the equation below (n = 1 diffraction order, d is the interplanar distance, λ is the wavelength and **θ** is the angle.)

$$2d\sin\theta = n\lambda \quad (1)$$

The lattice constant a, b and c of monoclinic structure of S1, S2, S3 and S4 are calculated using the equation 2 as shown below [26]

$$\frac{1}{d^2} = \frac{1}{\sin^2\beta}\left(\frac{h^2}{a^2} + \frac{k^2\sin^2\beta}{b^2} + \frac{l^2}{c^2} - \frac{2hl\cos\beta}{ac}\right) \quad (2)$$

The Debye Scherrer formula was used for the calculation of average crystallite size of predominant peaks which is given by in Eq. (3) [27],

$$D = \frac{k\lambda}{\beta\cos\theta} \quad (3)$$

Here, in above equation k is Scherrer constant and its value is 0.9, λ is the wavelength of Cu kα, $\beta$ is full-width at a half maximum (FWHM) and $\theta$ is the angle.

The average crystallite size of S1, S2, S3 and S4 were found to be 56.96, 46.59, 42.86 and 18.55 nm, respectively. Similar results of a reduction in particle size have been reported in literature due to the dual doping effect [28-30]. The lattice parameters of XRD like d-spacing and lattice constant a, b and c are listed in Table 1.

The results listed in Table 1 indicate that the lattice parameter changes after the incorporation of dopants into the host $WO_3$. This is because the $Cu^{2+}$ and $Ce^{4+}$ ions have higher ionic radii than the $W^{6+}$ ion. After doping with Ce and Cu, the crystallite size decreases abruptly due to the increased presence of oxygen vacancies and defect states in S4. The FTIR spectra of S1, S2, S3 and S4 indicate that Ce-Cu doping in $WO_3$ leads to a greater increase in the intensity of the 1217 cm$^{-1}$ and 1738 cm$^{-1}$ bands, which are associated with carbonate and carbonyl species. This suggests the formation of oxygen vacancies and surface defects, which may contribute to reduce particle growth and reduction in crystallite size.

**Table 1** XRD Analysis of S1, S2, S3 and S4

| Sample | Plane (hkl) | Peak Position (2$\theta$) | FWHM | Crystallite size (nm) | Average crystallite size (nm) | d value (Å) | Lattice Parameter (Å) |
|---|---|---|---|---|---|---|---|
| S1 | (002) | 23.17012 | 0.12 | 67.5791 | 56.9624 | 3.836041 | a = 7.2974 |
| | (020) | 23.65748 | 0.17683 | 45.9008 | | 3.754332 | b = 7.5086 |
| | (200) | 24.39284 | 0.14158 | 57.4073 | | 3.64835 | c = 7.6729 |
| S2 | (002) | 23.10402 | 0.149 | 54.4197 | 46.5917 | 3.84655 | a = 7.3142 |
| | (020) | 23.60806 | 0.18799 | 43.1720 | | 3.765553 | b = 7.5310 |
| | (200) | 24.32114 | 0.19265 | 42.1834 | | 3.65674 | c = 7.6939 |
| S3 | (002) | 23.18968 | 0.18922 | 42.8590 | 42.8676 | 3.83253 | a = 7.2838 |
| | (020) | 23.67822 | 0.19195 | 42.2868 | | 3.75455 | b = 7.5090 |
| | (200) | 24.42441 | 0.18704 | 43.4571 | | 3.64151 | c = 7.6659 |
| S4 | (002) | 23.14421 | 0.51347 | 15.7927 | 18.5550 | 3.83996 | a = 7.2990 |
| | (020) | 23.75107 | 0.50099 | 16.2039 | | 3.7432 | b = 7.4864 |
| | (200) | 24.37283 | 0.34339 | 23.6682 | | 3.6491 | c = 7.6807 |

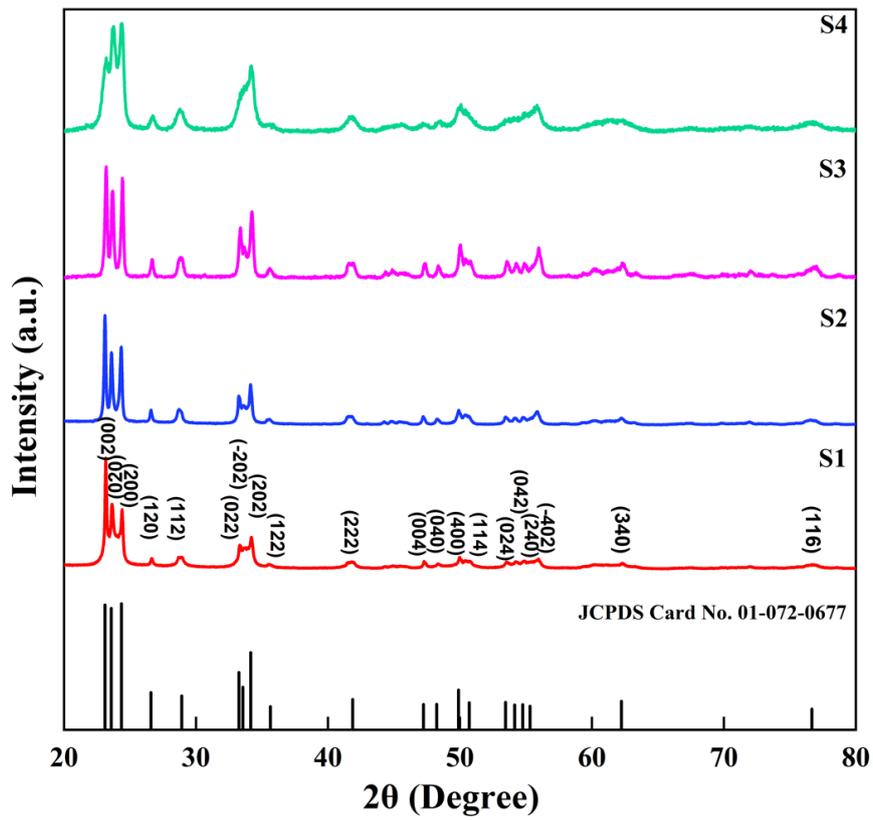

**Figure 2** XRD of S1 (Pure WO$_3$), S2 (Ce-doped WO$_3$), S3 (Cu-doped WO$_3$) and S4 (Ce-Cu-doped WO$_3$)

*3.1.2 Field Emission Electron Microscope (FE-SEM) Analysis:*

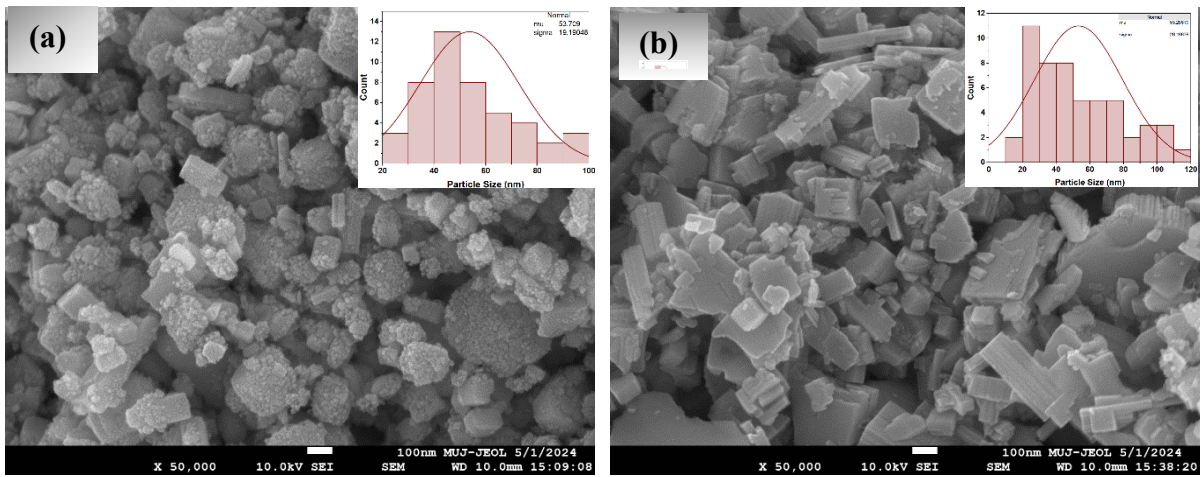

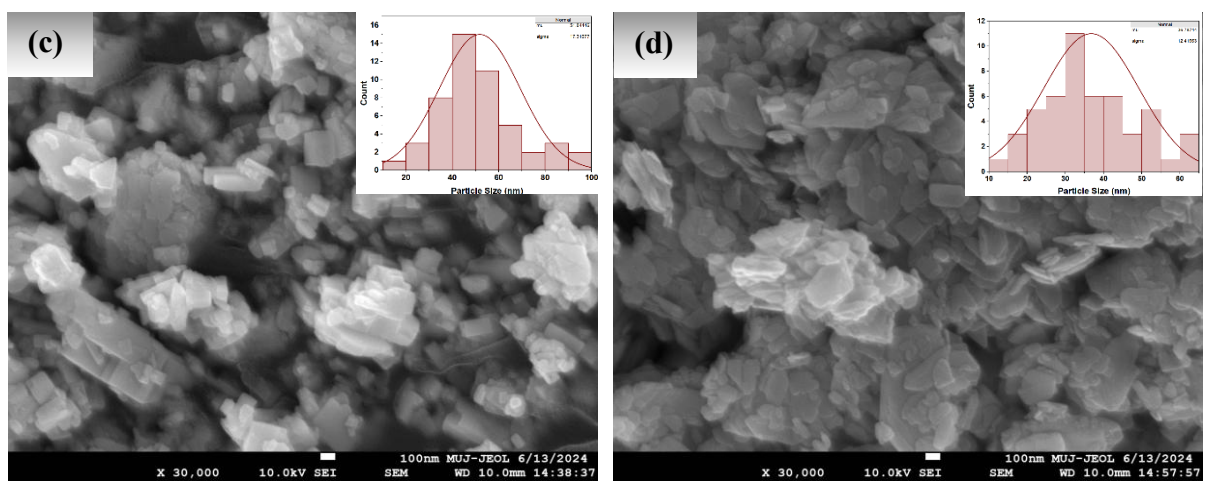

**Figure 3.** FESEM images of the (a) Pure $WO_3$ (S1) (b) Ce-doped $WO_3$ (S2) (c) Cu-doped $WO_3$ (S3) and (d) Ce-Cu-doped $WO_3$ (S4)

The morphology study of pure $WO_3$, Ce-doped $WO_3$, Cu-doped $WO_3$ and Ce-Cu-doped $WO_3$ nanocrystals was carried out by field scanning electron microscope (FESEM) and the FESEM images are shown in Figure 3. Figure 3 shows irregular shapes with some nanorods and plate-like structure which is caused by the capping effect of water-EG solvent [31]. After doping of Ce in $WO_3$, the morphology changed to regular rectangular plates as shown in Figure 3(b). This can be attributed to the incorporation of $Ce^{3+}$ in the host $WO_3$ lattice. After doping of Cu in $WO_3$, the morphology changes to rectangular plates but shows agglomeration. Doping of Ce and Cu in $WO_3$ results in a change in morphology and irregular plates with reduced thickness appear. FESEM images of Figure 3 were analysed by ImageJ software to calculate the average grain size. The average width of S1, S2, S3 and S4 nanocrystals were obtained to be 53.709 nm, 53.28 nm, 51.84 nm and 36.78 nm.

*3.1.3 EDAX Analysis*

The purity of all samples was confirmed by EDX analysis. The elemental mapping of W, O, Ce and Cu are shown in Figure 4. The results indicated that even distribution of elements shows successful doping of Ce and Cu.

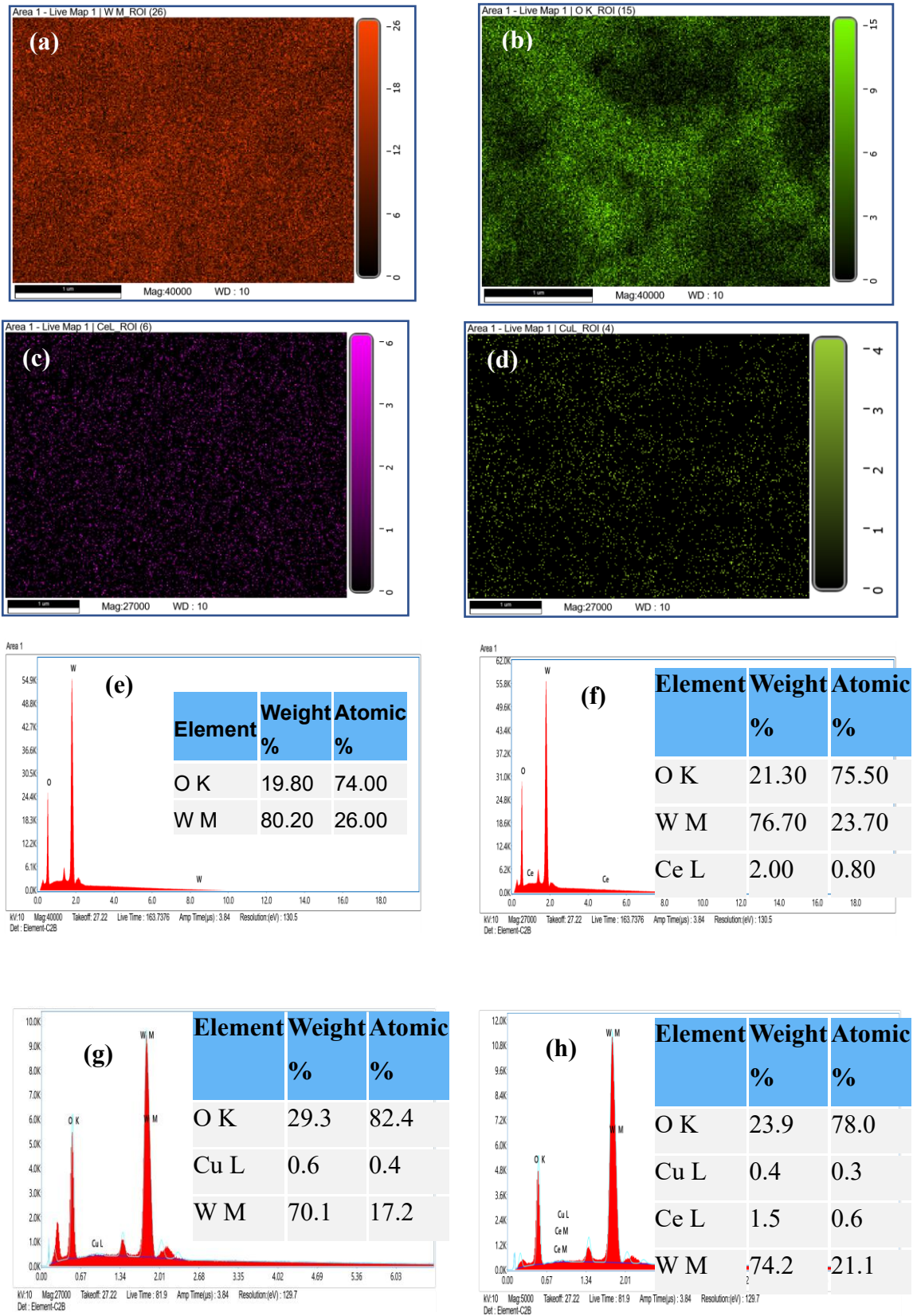

**Figure 4** EDAX Colour mapping for (a) Tungsten (b) Oxygen (c) Cerium and (d) Copper; (e),(f), (g) and (h) are the elemental spectrum and inset images represent atomic weight percent of elements.

*3.2 Optical Characterization*

*3.2.1 PL (Photoluminescence Spectroscopy)*

PL spectroscopy is a useful method that can help to reveal the energy band structures, surface oxygen vacancies, defects and impurity levels in the material, as well as the immigration and transfer of electrons/holes in the nanomaterial [32]. The PL spectra of these samples are shown in Figure 5. The PL emission spectra of S1, S2, S3 and S4 show blue, green and red emission bands in the wavelength range of 300–750 nm. Five prominent peaks are observed in the PL spectra for S1, S2, S3 and S4 at wavelengths of 464, 479, 489, 542 and 695 nm. The increase in PL intensity in the green and red emission bands for S4 indicates higher oxygen vacancies and defect conditions [33]. The blue (420–480 nm), green (500–550 nm) and red (620–750) emission bands arise through electronic transitions facilitated by oxygen vacancies or defect levels, which lie below the conduction band minimum within the fundamental bandgap. [34]. These oxygen vacancies may be neutral, singly charged and doubly charged ($V_O^0$, $V_O^+$ and $V_O^{++}$) [32]. It is possible to trace the blue emission band to the band−band transition[35]. Regarding the transitions corresponding to defect levels, the green emission is produced by the transition from $V_O^+$ to $V_O^0$, while the red emission is produced by the transitions from $V_O^{++}$ to $V_O^+$ and $V_O^{++}$ to $V_O^0$, respectively [36]. The similar peaks are observed for the Ce-doped, Cu-doped and Ce-Cu-doped $WO_3$, respectively. The Figure 5 indicates that after doping PL intensity increases that indicates increase in concentration of oxygen vacancies which in turn enhances its gas sensing performance [37]. The Ce-doped $WO_3$ (S2) has maximum PL intensity due to the enhanced radiative recombination and maximum oxygen vacancies [22]. Cu-doped $WO_3$ (S3) has lower PL intensity from the S2 due to the quenching effect of Cu doping [12]. Copper introduces new defect sites and acts as non-radiative recombination centres. Instead of radiative recombination (PL emission), electrons are trapped or transferred, reducing the PL intensity. Moderate quenching improves gas sensing by enhancing the charge carrier interaction with gas molecules. Excessive quenching may reduce the effective charge carrier mobility and hinder overall conductivity changes, limiting gas sensitivity. The optical transitions involving $Ce^{3+}$ ions are often efficient, with high probabilities for photon emission due to the well-known f-d transitions in rare-earth ions. This makes $Ce^{3+}$ an effective dopant for enhancing PL in various host materials, including $WO_3$. The transitions involving Cu may involve d-d transitions, which are often less radiatively efficient compared to the f-d transitions in rare-earth elements like Ce [38]. Cu-Ce co-doping results in a decrease in PL intensity compared to Ce doping alone, indicating that Cu's quenching effect is stronger than Ce's enhancing effect.

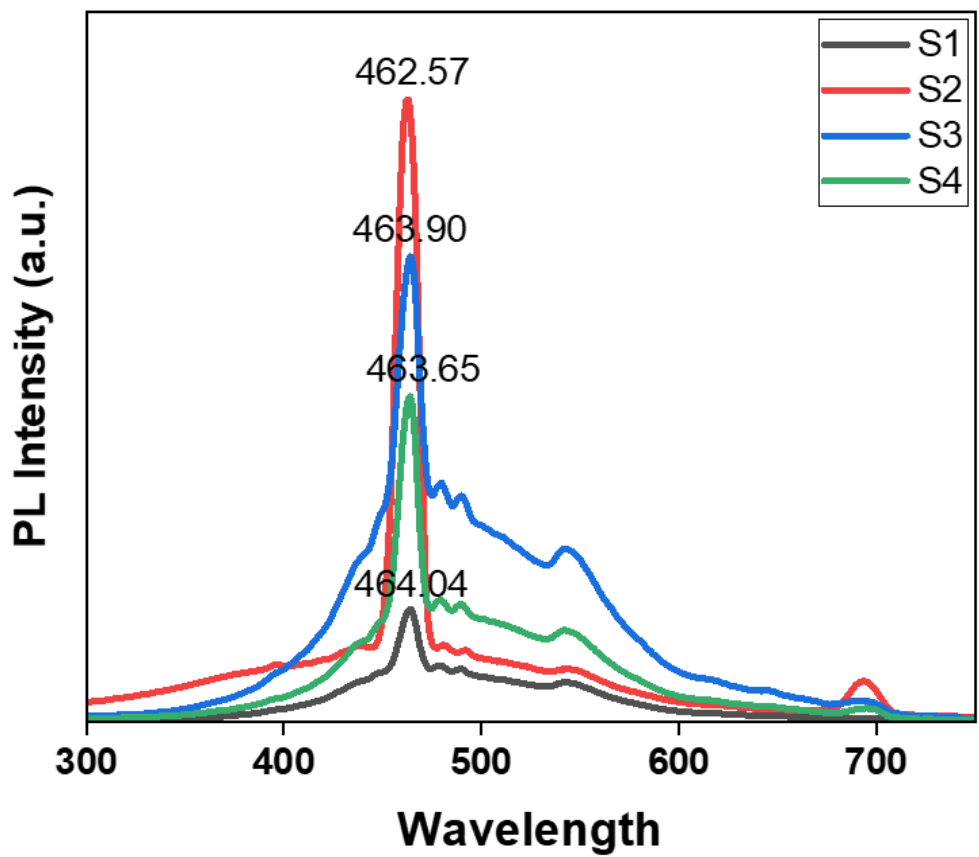

**Figure 5** Photoluminescence (PL) Spectra of S1, S2, S3 and S4

*3.2.2 UV-Visible Spectra*

The inset image of Figure 6 shows the absorption spectra of S1, S2, S3 and S4. This indicates that the S1, S2, S3 and S4 shows fairly identical absorption spectra in the region of 250-800 nm. The maximum absorbance of the samples occurs in the UV-region (< 300 nm). In the visible region the maximum absorption is observed in S2 (Ce-doped $WO_3$) which is due to the strong interaction between Ce and W because of 4f-5d interaction. The Ce-Cu-doped $WO_3$ has higher value of optical absorbance in the UV region that is due to the smaller crystallite size of Ce-Cu-doped $WO_3$. The electronic transitions including the conduction band, valence band, and different inherent defect levels are typically attributed to the optical absorption of nanocrystals [39]. The optical band gap energy of the samples was calculated by the Tauc equation. The Tauc equation relates the photon energy to the optical band gap energy of the semiconductor by the following equation [40]:

$$\alpha h\vartheta = A(h\vartheta - E_g)^n$$

where, A is the constant, $\alpha$ is the absorption coefficient, $h\vartheta$ is the energy of incident photon and $E_g$ is the optical band gap of the material. The exponent $n$ represent the nature of optical transition, for allowed indirect and direct optical transition it has value 2 and 0.5, respectively. The Tauc plot is represented in Figure 6. S1, S2, S3 and S4 have indirect band gap which results in lower recombination rate and increases the stability to the sensor. The band gap values of pure $WO_3$, Ce-doped $WO_3$, Cu-doped $WO_3$ and Ce-Cu-doped $WO_3$ are 2.62 eV, 2.48 eV, 2.65 eV and 2.64 eV, respectively. The Ce atom has partially filled 4f orbitals which creates additional energy states that reduce the band gap of $WO_3$ and also contributes to more oxygen vacancies [23]. The Cu incorporation in $WO_3$ induce lattice distortions that may increase the band gap value. The band gap value of the S4 is higher than the S2 but lower than the S3. This indicates that the synergistic effect of Ce and Cu.

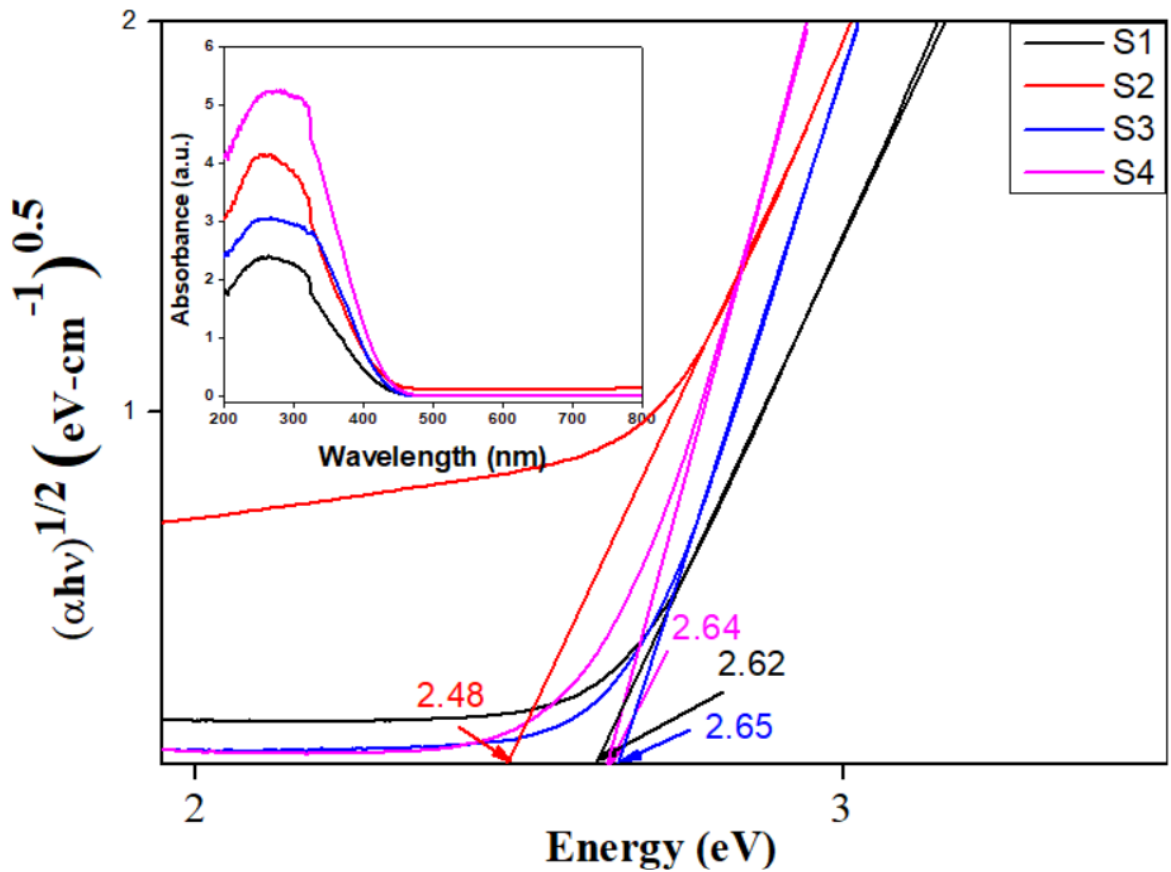

**Figure 6** Tauc Plot of S1, S2, S3 and S4, inset plot represent UV-absorption spectra of samples

*3.2.3 FTIR Spectroscopy*

Fourier transform Infrared (FTIR) spectroscopy was used to analyse the chemical vibrational modes on the surface of nanocrystals. It was also used to determine the structural and chemical behaviour of the nanocrystals. The FTIR spectra of S1, S2, S3 and S4 are shown in Figure 7. The pure and doped samples have the same IR active regions. The absorption bands that occur in the fingerprint region at 726, 733, 723, 714 817, and 819 cm$^{-1}$ are due to the stretching and bending vibrations of O-W-O and W-O-W [41]. After doping, the shifting and broadening of these peaks (733-723 cm$^{-1}$) indicate the structural modification which enhance gas sensing property. This mode of vibrations represents the monoclinic phase of tungsten oxide [42]. The bands at 1217 and 1738 cm$^{-1}$ are due to the C-O stretching in the carbonyl group which occurs because of ethylene glycol that was used in the sample preparation [43]. The peak intensities at 1217 and 1738 cm$^{-1}$ in S4 suggest the presence of surface defects and oxygen vacancies which are beneficial for gas sensing. The peak at 1526 cm$^{−1}$ is due to the vibration of tungsten-hydroxyl (W-OH) bond [44]. The low intensity peak observed at 2348 cm-1 is the absorption of $CO_2$ molecule from the atmosphere on the surface of the material. The bands appearing above 3000 cm$^{-1}$ are due to the stretching vibration of O-H group. The low intensity of these peaks indicates less water absorption from the atmosphere for all samples [42]. The low intensity of these peaks in S2, S3 and S4 as compared to S1 indicates less -OH group and less water absorption. This property is beneficial for humidity independent gas sensing studies.

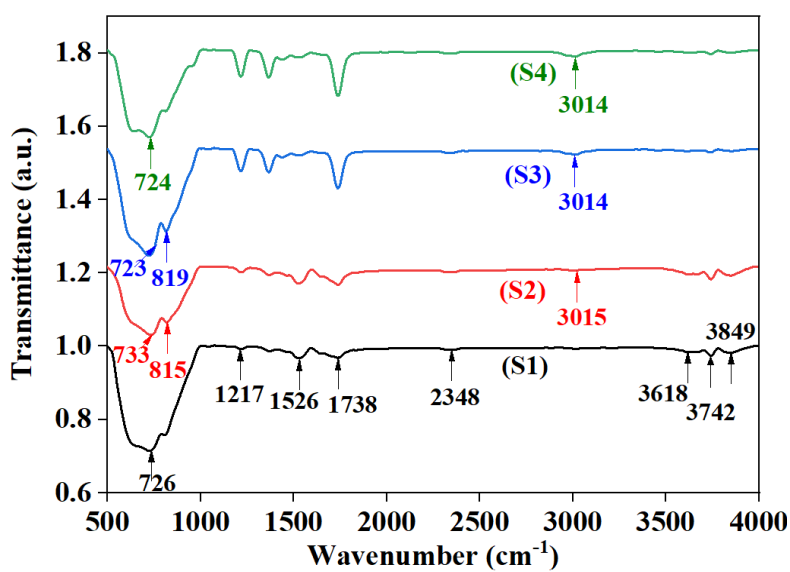

**Figure 7** FTIR spectra of S1, S2, S3 and S4

## 3.3 Gas Sensing Measurement

### 3.3.1 Fabrication of gas sensing devices

Sensing devices were made up of the screen-printing method. The alumina microtubes were used to screen printing. The screen-printed alumina tubes were mounted on four electrodes as shown in figure 8. The four electrodes made up of gold wires were soldered on the four contacts of the pedestal. An agate mortar grinder was used to mark and grind all test samples into fine particles. Absolute ethanol was then added, and the grounded samples were gently printed on the alumina. After, heat treatment of 1hour the screen-printed sample tubes were ready for gas sensing measurements.

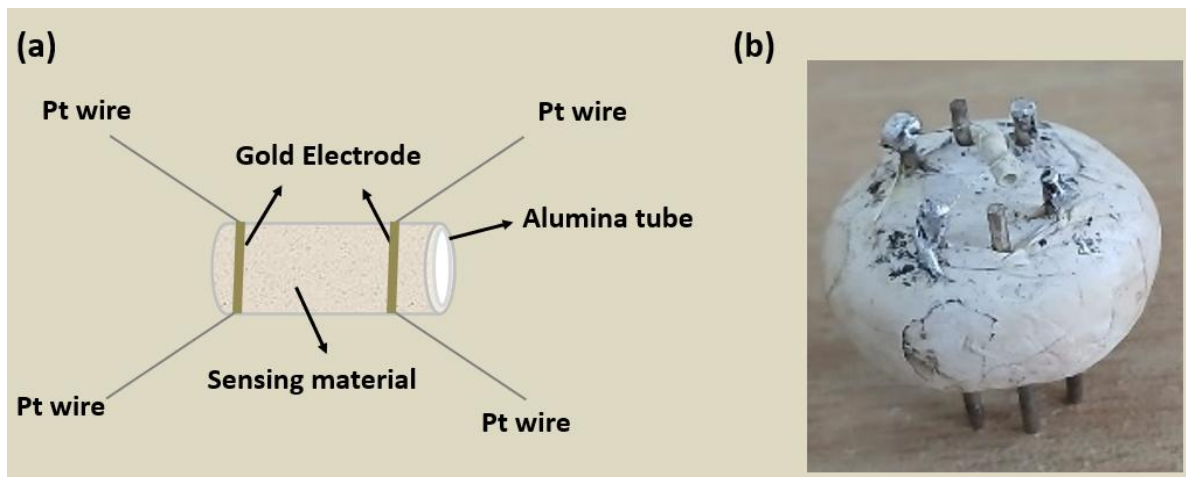

**Figure 8 (a)** Schematic diagram of tubular sensor (b) Real image of fabricated sensor on the pedestal

For gas-sensing measurements, the coated sample tubes were tested in a chamber under controlled temperature, pressure, and gas composition. The impedance measurement was carried out by IM 3536 LCR meter, in the frequency range 400 Hz-4 MHz in the temperature range of 130°C-290°C. The electrical behaviour of the samples was analysed by the impedance plots. The curves were obtained at different temperatures and for different gases of different categories such as methanol, ammonia, toluene and formaldehyde [45].

### 3.3.2 Temperature dependent Impedance behaviour of Samples in dry Air

Cole-Cole plots are constructed for the temperature in the range of 130-290°C for the interval of 40°C in dry air. In all AC measurement, we use impedance (combination of resistance and reactance) instead of purely resistive network. The Cole-Cole plot is a plot between ZCosθ (Z') and ZSinθ (Z''), where θ is the phase angle and Z is the impedance. The lower temperature values (130°C and 170°C), impedance study is discarded due to smaller changes in impedance. After doping Cu and Ce into $WO_3$ (S4), the conductive properties improved, allowing us to obtain the impedance plot for gas sensing at 170°C. The impedance plots for the temperature range 210-290°C for S1, S2, S3 and S4 are shown in Figure 9. The impedance of the S1, S2, S3 and S4, decreases with increase in temperature throughout the temperature range. After doping of Ce in $WO_3$, the impedance increases for whole temperature range and impedance decreases monotonously with increasing temperature. S3 shows a smaller diameter in the Nyquist plot compared to pure $WO_3$ across all temperatures and it decreases as temperature rises. S4 exhibits a smaller impedance value than S1 within the temperature range of 210°C to 290°C. At 210°C, S4 shows a lower impedance value compared to S2 and S3. However, at 250°C and 290°C, its impedance is higher than that of S3 but lower than that of S2. The impedance value increases for S2 than the s (S1, S3 and S4) because Ce-doping can make the grains in the material smaller. Smaller grains mean there are more grain boundaries, which block the flow of charge and increase impedance [46]. The lower impedance value of S3 compared to S1 indicates that Cu doping may reduce the density of charge-trapping defects or vacancies that act as barriers to charge transport. This allows carriers to move more freely through the material, resulting in decreasing impedance [21]. For S4, the impedance value is lower compared to S1, S2, and S3 at 210°C, because the combination of Ce and Cu doping can synergistically introduce additional free charge carriers into the $WO_3$ matrix. At 210°C, these carriers may contribute to better conductivity compared to Cu-doped $WO_3$ alone, leading to reduced impedance. But at 250°C and 290°C, the impedance value is more for S3 because Ce doping may enhance defect-related effects at higher temperatures. Oxygen vacancies can act as recombination centres or as barriers to carrier mobility. The S3 and S4 samples have two semicircles as shown in Figure 12 and 13.

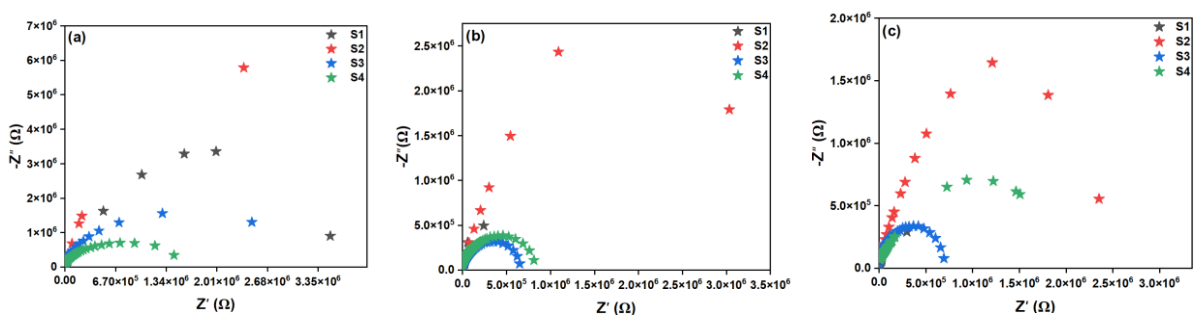

**Figure 9** Temperature dependent Impedance Plot (Cole-Cole plot) of S1, S2, S3 and S4 at (a) 210°C, (b) 250°C, (c) 290°C temperature

### 3.3.3 Impedance behaviour of S1, S2, S3 and S4 after exposure to gases:

The present study focuses on the detection of methanol, ammonia, toluene, and formaldehyde (HCHO) due to their widespread usage and associated risks. From these, ammonia is a commonly used in manufacturing and household products. Also, ammonia is a major indoor air pollutant, signifying the importance of its monitoring. The present research work is based on the detection of ammonia gas. For this study the concentration of gases was fixed at 400 ppm. The response or sensitivity was calculated by the formula [47]

$$S = \frac{|Z_a|}{|Z_g|} \qquad (4)$$

where, $|Z_a|$ is the magnitude of total impedance in the presence of air and $|Z_g|$ is the magnitude of total impedance in the presence of gases.

At 210°C, the Cole-Cole plot exhibits unusual behaviour for S1 due to a reduced relaxation process at this temperature. Hence we exclude S1 at 210°C for analysis. For further studies, we investigated the samples at 210°C (S2, S3 and S4), 250°C, and 290°C. At 250°C, ammonia, methanol, and toluene exhibit reducing behaviour, as indicated by a decrease in the impedance plot upon exposure to these gases. However, formaldehyde (HCHO), which is also a reducing gas, shows a higher impedance value compared to air. This anomaly could be attributed to the formation of intermediates like formic acid during the sensing mechanism. Additionally, lower selectivity is observed at this temperature, as shown in the figure 10(a). At 290°C, ammonia gas displays a lower diameter value of semicircular Cole-Cole plot compared to all other gases, indicating good selectivity and response toward ammonia (shown in Figure 10(b)). While HCHO and methanol demonstrate medium response value at this temperature, toluene shows poor selectivity.

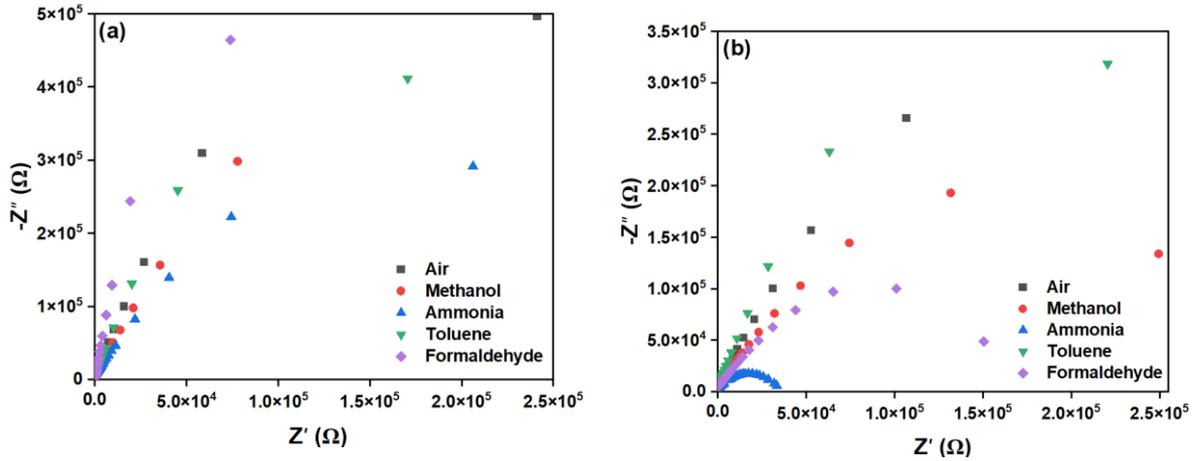

**Figure 10** Nyquist plot of S1 for different gases at 400 ppm concentration (a) at 250°C, (b) 290°C

After doping $WO_3$ with cerium (Ce), the Cole-Cole plot at 210°C indicates good selectivity for methanol gas, as evident from the smaller semi-circular radius as shown in Figure 11(a). Among the other gases, methanol shows the maximum response, while toluene and HCHO exhibit lower responses. At 250°C, ammonia gas displays a smaller semicircular arc compared to methanol and toluene (shown in Figure 11(b)). While HCHO shows a higher impedance value than air, likely due to the formation of intermediates such as formic acid, which can trap electrons. At 290°C, both ammonia and methanol exhibit reducing behaviour, with ammonia displaying the smallest semicircular radius on the Cole-Cole plot. Conversely, at this temperature, toluene and HCHO exhibit oxidizing behaviour, leading to an increase in the radius of the semicircular impedance plot due to the formation of intermediates.

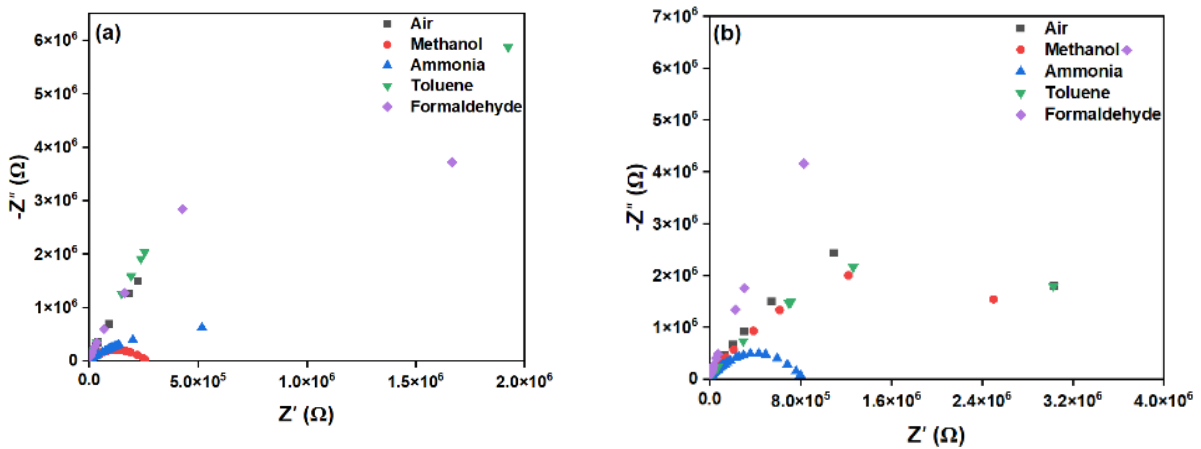

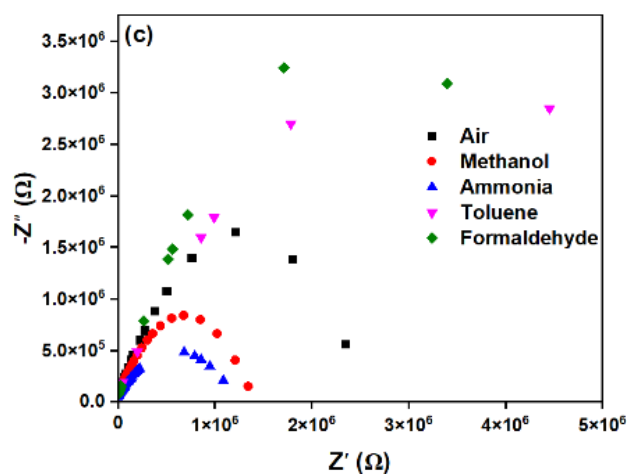

**Figure 11** Nyquist plot of S2 for different gases at 400 ppm concentration (a) at 210°C, (b) 250°C, (c) 290°C

The gas sensing behaviour of Cu-doped $WO_3$ for methanol, ammonia, toluene, and formaldehyde (HCHO) was analysed using Cole-Cole plots shown in Figure 12. The Nyquist plot exhibits two semicircles, it generally indicates the presence of two different relaxation processes in the system. The first semicircle represents the bulk grain response at higher frequency region[48]. The second semicircle represents the grain boundary resistance and surface related phenomena at lower frequency region [46]. At lower temperatures, the grain boundary resistance ($R_{gb}$) is dominant, leading to a larger semicircle at lower frequencies. At higher temperatures, grain conductivity improves, and both semicircles shrink, indicating reduced total resistance. At 210°C, ammonia and HCHO exhibited smaller semicircular arcs in the plot, indicating higher responses compared to toluene and methanol. However, at this temperature, ammonia and HCHO demonstrated poor selectivity, as shown in Figure 12(a). At 250°C, methanol, ammonia, and HCHO displayed reducing behaviour, as evidenced by smaller semicircular arcs. Among these, methanol showed the highest response but only moderate selectivity. Toluene exhibited a higher impedance value compared to air, likely due to the formation of intermediates during the sensing process. At 290°C, ammonia demonstrated the highest response value with good selectivity relative to toluene and HCHO but only moderate selectivity compared to methanol. Formaldehyde exhibited a higher impedance value than air, potentially caused by the formation of intermediates.

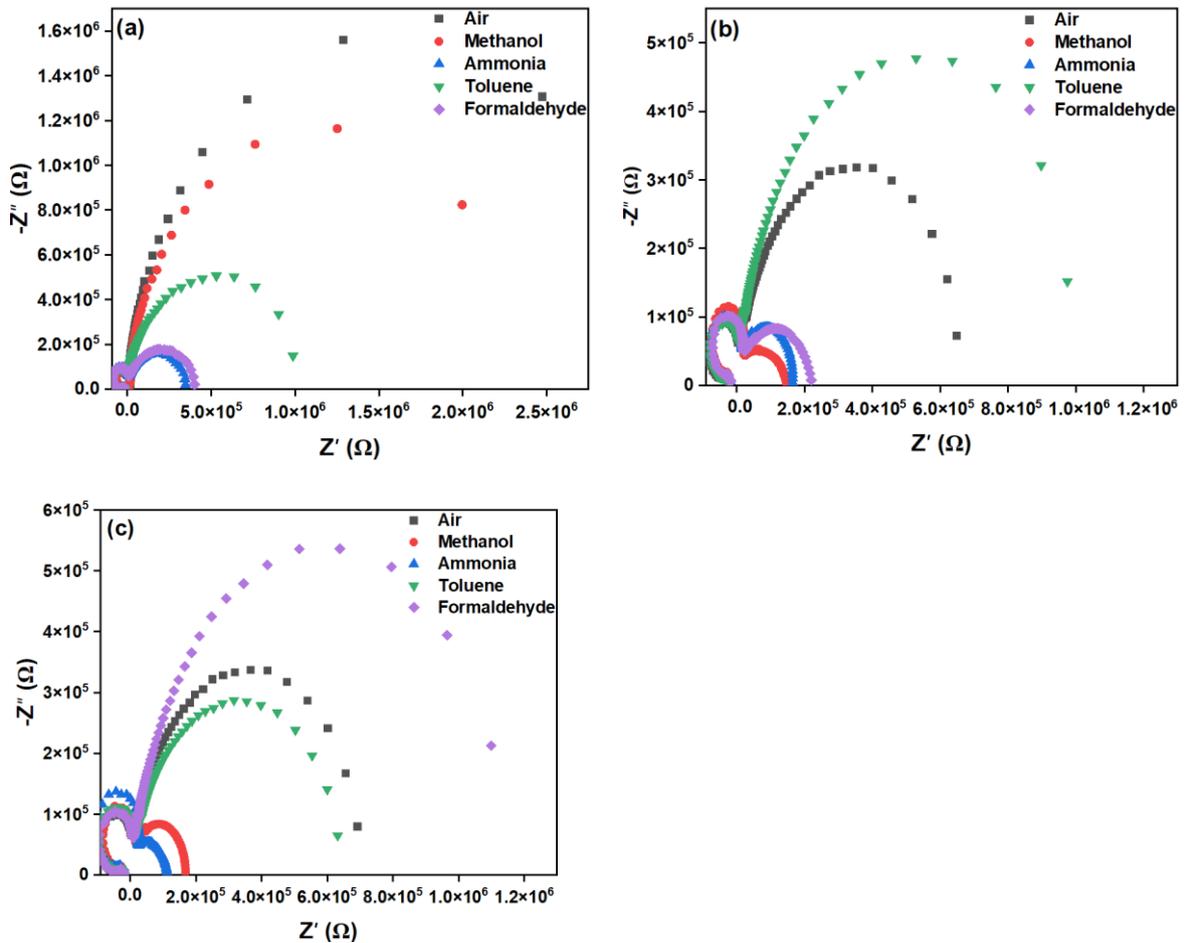

**Figure 12** Nyquist plot of S3 for different gases at 400 ppm concentration (a) at 210°C, (b) 250°C, (c) 290°C

The gas sensing behaviour of Ce-Cu-doped $WO_3$ is shown in Figure 13. The two semicircles suggest bulk (high-frequency) and grain boundary (low-frequency) contributions. This sample exhibited response values at 170°C for all tested gases, but the selectivity was poor. At 210°C, ammonia displayed a smaller semicircular arc in the Cole-Cole plot, indicating reducing behaviour, while HCHO exhibited a higher impedance value compared to air. At 250°C, ammonia showed a smaller semicircular arc and better selectivity. Methanol and toluene also demonstrated reducing behaviour, while HCHO exhibited a higher impedance value than air. At 290°C, all gases—ammonia, HCHO, methanol, and toluene—exhibited reducing behaviour. Ammonia, HCHO, and toluene showed smaller semicircular arcs but poor selectivity. Methanol exhibited moderate selectivity at this temperature.

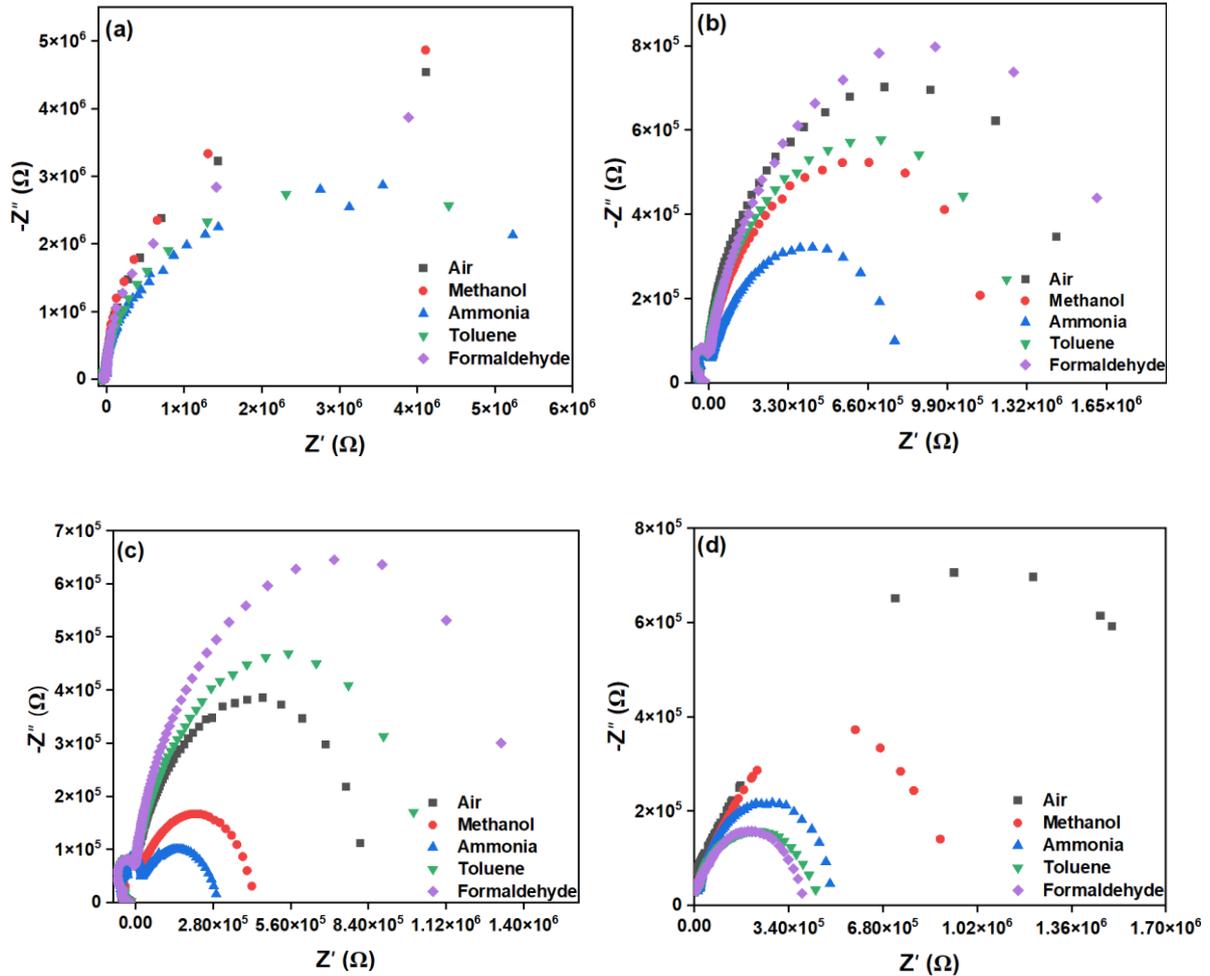

**Figure 13** Nyquist plot of S4 for different gases at 400 ppm concentration (a) at 170°C, (b) 210°C, (c) 250°C and (d) 290°C

*3.3.4 Frequency dependent sensitivity of gases at different temperature:*

Figure 14 shows the gas sensing response as a function of logarithm of frequency for methanol, ammonia, toluene, and formaldehyde at different temperatures. For S1 at 250°C, the response is highest for toluene, ammonia, and formaldehyde at ~1.07 MHz. However methanol has a similar response at all frequencies. At 290°C, ammonia shows higher sensitivity at lower frequencies (log f < 5). Between log f ~ 5.0 and 6.0, all gases exhibit similar sensitivity, reducing selectivity. This suggests S1 has higher sensitivity at specific frequencies.

For S2 at 250°C, ammonia shows maximum sensitivity (~4.5) at lower frequencies. Selectivity decreases in the mid-frequency range and increases at higher frequencies. At 290°C, toluene and formaldehyde exhibit higher sensitivity at 0.13 MHz due to their higher impedance in the Cole-Cole plot. Ammonia sensitivity gradually decreases across all frequencies.

For S3 at 250°C, methanol and ammonia show higher sensitivity at lower frequencies. Sensitivity decreases with frequency, reducing selectivity. Toluene response increases at higher frequencies. At 290°C, ammonia has the highest response (~6.5) at low frequencies, followed by methanol (~4.0). The response of toluene and formaldehyde remain nearly constant (~1.0) over the entire frequency rang. At higher frequencies (log f > 5.5), all gases converge to ~1.0 response.

For S4, sensing response and selectivity are observed at lower temperatures. At 210°C, ammonia sensitivity decreases, but selectivity improves. At 250°C, ammonia sensitivity increases, making S4 a thermally stable sensor for ammonia. These results highlight how Ce-Cu doping in $WO_3$ enhances ammonia sensing stability. Additionally, S4 (Ce-Cu-doped $WO_3$) exhibits higher sensitivity toward ammonia compared to other gases at 170°C, 210°C, and 250°C, further confirming its thermal stability as an ammonia sensor.

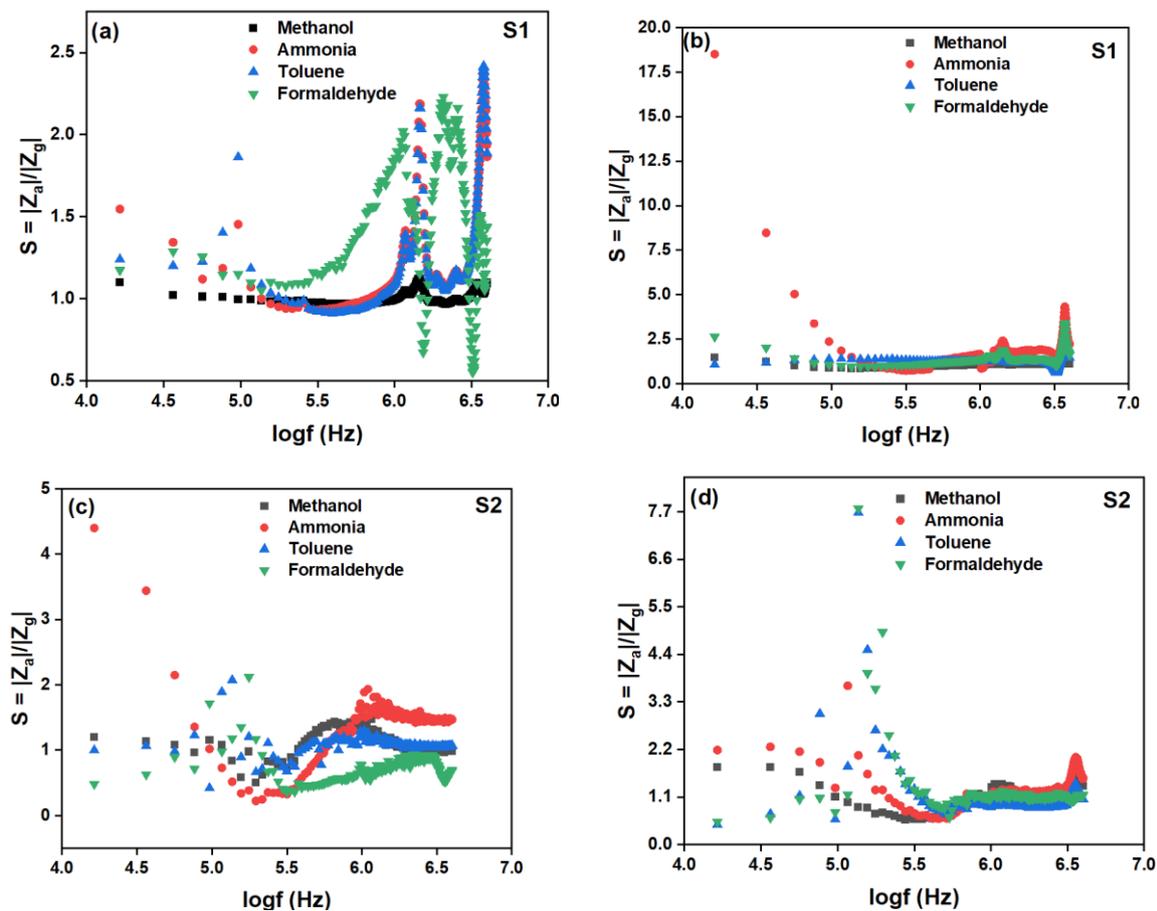

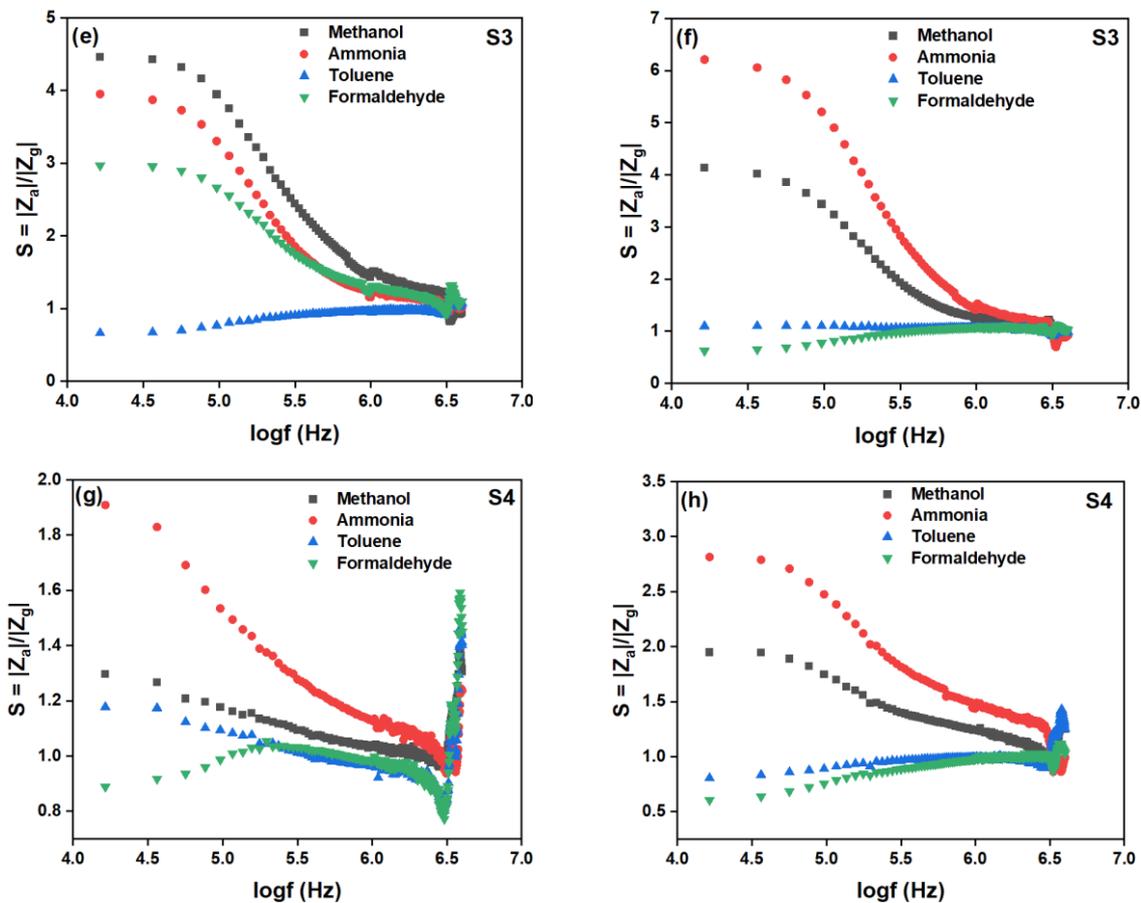

**Figure 14** Frequency dependent sensitivity behaviour of different gases at (a) and (b) for S1 at 250°C and 290°C, respectively ;(c) and (d) for S2 at 250°C and 290°C, respectively ;(e) and (f) for S3 at 250°C and 290°C, respectively; (g) and (h) for S4 at 210°C and 250°C

### 4. Gas sensing mechanism

For pristine $WO_3$, the effect of gas molecules on the surface is seen because of the modification in the space charge region that is formed on the surface of the sensing layer. The space charge region is created by the adsorption of oxygen species present in the environment. Although, for doped (Ce, Cu and Ce-Cu) $WO_3$ response is observed from the adsorption of oxygen species but significant contribution come from the dopant atoms. These dopant atoms modify the depletion region and amplify the sensing material properties. The ammonia gas exhibits reducing behaviour; due to this behaviour, electrons flow from the gas molecule to the sensing layer. That reduces the potential barrier height at the interface. The decrease in band bending after the exposure of gas molecules is due to the decrease in barrier potential after the doping of dopants shown in the Figure 15. It is known from the literature [49] that the stable oxygen ions were $O_2^-$ at temperatures below 100°C. Those oxygen molecules change to $O^-$ between

100 and 300° C, and the same oxygen molecules change to $O^{2-}$ above 300°C. Since we have set the operating temperature at 250°C, which is ideal, the following reaction scheme is applicable. As depicted in Figure 15, when $WO_3$ exposed to air, oxygen molecules from the environments adsorb onto its surface. These molecules capture electrons from the conduction band and form the adsorbed oxygen ions such as $O^-$, $O_2^-$ and $O^{2-}$. This process creates a depletion layer, a region with fewer charge carriers. Once the depletion region stabilizes, the sensor remains in the high-resistance state until the external factors (such as heat and gas molecules) alter the oxygen adsorption dynamics. When $NH_3$ gas molecules are introduced, the adsorbed oxygen $O^-$ ions react with them, converting into $N_2$, $H_2O$ and releasing electrons on the material's surface. As a result, electron-depleted region narrows. This leads to reduction in the sensor's resistance as depicted in the Cole-Cole plot profile. The Ce-Cu-doped $WO_3$ exhibit good $NH_3$ sensing capability due to their modified effect of Ce-Cu, which introduce additional defect states and oxygen vacancies. Furthermore, as illustrated in the band diagram mechanism in Figure 15, the lower work function of Ce (2.7 eV) and Cu (4.65 eV) relative to the $WO_3$ work function (5.05 eV), facilitates the charge migration [50]. The difference in work functions cause electron to flow between energy bands, leading to band bending. As a result, the Fermi level shifts to a new equilibrium, forming a depletion region and a Schottky barrier at the interfaces [51]. These changes affect the concentration and movement of charge carriers, which in turn modify the baseline resistance and enhance the sensor's properties in terms of sensitivity, selectivity and stability [52].

$$2NH_3 + 3O^- \rightarrow N_2 + 3H_2O + 3e^-$$

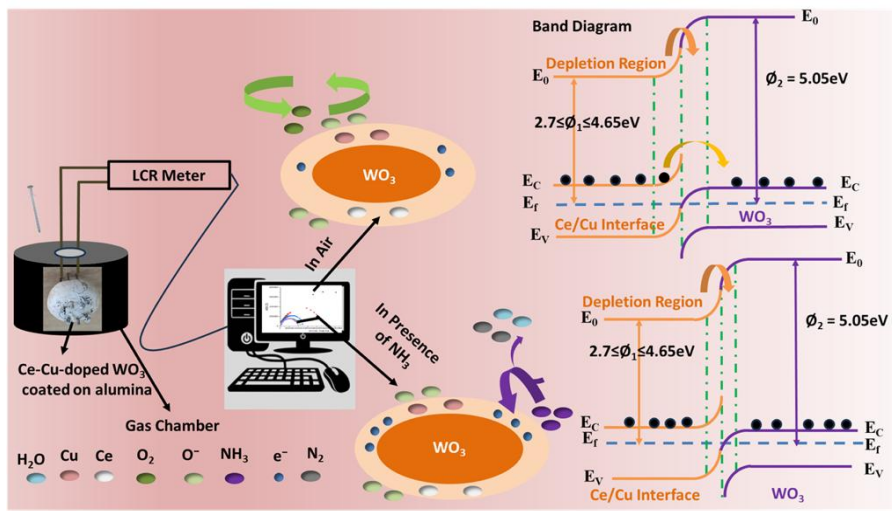

**Figure 15** Schematic diagram of ammonia sensing mechanism and band structure diagram

## 5. Conclusion

This study investigated the structural, optical and gas sensing properties of $WO_3$, Ce-$WO_3$, Cu-$WO_3$ and Ce-Cu-doped $WO_3$ experimental approach. The samples were synthesized using the hydrothermal method. FESEM images showed that the samples exhibited rectangular plate structures that were modified after doping with Ce, Cu and Ce/Cu. The average crystallite sizes of S1, S2, S3 and S4 were 56.96, 46.59, 42.86 and 18.55 nm, respectively. XRD results confirmed the successful doping, while UV-analysis showed modifications in the band gap between 2.48-2.65 eV. In addition, FTIR was conducted to identify the functional groups. Additionally, PL analysis indicated that the green and red emission bands (542 and 695 nm) of S4 increased in PL intensity, suggesting an increase in the formation of oxygen vacancies and defect states. The optimum temperature for gas sensing studies was determined to be 250 °C. Impedance-based gas sensing experiments demonstrated that Ce-Cu-doped $WO_3$ exhibits high selectivity and thermal stability for ammonia detection, performing better than undoped $WO_3$. The sensor showed better response at low operating temperatures, making it an energy-efficient candidate for practical applications.

6. **Acknowledgement**
7. The authors acknowledge the DST PURSE grant no. SR/PURSE/2021/77 awarded to Amity University Rajasthan, Jaipur, India by Govt of India for the financial support. **SK acknowledges the financial support from Science and Engineering Research Board (SERB), Department of Science and Technology (DST), India** (grant No. EEQ/2023/000676) and IIT Jodhpur for a research initiation grant (grant no. I/RIG/SNT/20240068).

## References

1. Jagannathan, M., et al., *Selective room temperature ammonia gas sensor using nanostructured ZnO/CuO@ graphene on paper substrate.* Sensors and Actuators B: Chemical, 2022. **350**: p. 130833.
2. Cao, Y., et al., *Tungsten oxide clusters decorated ultrathin In2O3 nanosheets for selective detecting formaldehyde.* Sensors and Actuators B: Chemical, 2017. **252**: p. 232-238.
3. Liu, X., et al., *A survey on gas sensing technology.* Sensors, 2012. **12**(7): p. 9635-9665.
4. Simchi, H., et al., *Structural, optical, and surface properties of WO3 thin films for solar cells.* Journal of alloys and compounds, 2014. **617**: p. 609-615.
5. Sharmila, B. and P. Dwivedi, *Optical sensing and computing memory devices using nanostructured WO3.* Materials Science in Semiconductor Processing, 2024. **173**: p. 108106.

6. Yadav, P.K., et al., *Enhanced performance of WO3 photodetectors through hybrid graphene-layer integration.* ACS Applied Electronic Materials, 2021. **3**(5): p. 2056-2066.
7. Dong, C., et al., *A review on WO3 based gas sensors: Morphology control and enhanced sensing properties.* Journal of Alloys and Compounds, 2020. **820**: p. 153194.
8. Zhang, R., et al., *Oxygen vacancy engineering of WO3 toward largely enhanced photoelectrochemical water splitting.* Electrochimica Acta, 2018. **274**: p. 217-223.
9. Yagi, M., et al., *Preparation and photoelectrocatalytic activity of a nano-structured WO3 platelet film.* Journal of Solid State Chemistry, 2008. **181**(1): p. 175-182.
10. Marnadu, R., et al., *Impact of phase transformation in WO3 thin films at higher temperature and its compelling interfacial role in Cu/WO3/p–Si structured Schottky barrier diodes.* Zeitschrift für Physikalische Chemie, 2020. **234**(2): p. 355-379.
11. Efkere, H.İ., et al., *Investigation of the effect of annealing on the structural, morphological and optical properties of RF sputtered WO3 nanostructure.* Physica B: Condensed Matter, 2021. **622**: p. 413350.
12. Deepa, B. and V. Rajendran, *Pure and Cu metal doped WO3 prepared via co-precipitation method and studies on their structural, morphological, electrochemical and optical properties.* Nano-structures & Nano-objects, 2018. **16**: p. 185-192.
13. Sriram, S.R., et al., *Synthesis and characterization of pure and Cu-doped WO3 thin films for high performance of toxic gas sensing applications.* Applied Surface Science Advances, 2023. **15**: p. 100411.
14. Wang, Y., et al., *Construction of doped-rare earth (Ce, Eu, Sm, Gd) WO3 porous nanofilm for superior electrochromic and energy storage windows.* Electrochimica Acta, 2022. **412**: p. 140099.
15. Govindaraj, T., et al., *The remarkably enhanced visible-light-photocatalytic activity of hydrothermally synthesized WO3 nanorods: an effect of Gd doping.* Ceramics International, 2021. **47**(3): p. 4267-4278.
16. Mohanraj, M., et al., *Investigation on the microstructural, optical, electrical, and photocatalytic properties of WO3 nanoparticles: An effect of Ce doping concentrations.* Journal of Materials Science: Materials in Electronics, 2023. **34**(28): p. 1961.
17. Liu, Y., et al., *Engineering pore walls of mesoporous tungsten oxides via Ce doping for the development of high-performance smart gas sensors.* Chemistry of Materials, 2022. **34**(5): p. 2321-2332.
18. Tran, V.A., et al., *Excellent photocatalytic activity of ternary Ag@ WO3@ rGO nanocomposites under solar simulation irradiation.* Journal of Science: Advanced Materials and Devices, 2021. **6**(1): p. 108-117.
19. Mohammadi, S., et al., *Preparation and characterization of zinc and copper co-doped WO3 nanoparticles: application in photocatalysis and photobiology.* Journal of Photochemistry and Photobiology B: Biology, 2016. **161**: p. 217-221.
20. Mehmood, F., et al., *Facile synthesis of 2-D Cu doped WO3 nanoplates with structural, optical and differential anti cancer characteristics.* Physica E: Low-Dimensional Systems and Nanostructures, 2017. **88**: p. 188-193.
21. Dong, X., et al., *Efficient charge transfer over Cu-doped hexagonal WO3 nanocomposites for rapid photochromic response.* Journal of Photochemistry and Photobiology A: Chemistry, 2022. **425**: p. 113716.
22. Haroon, A., K. Anwar, and A.S. Ahmed, *Visible light-driven photo remediation of hazardous cationic dye via Ce-doped WO3 nanostructures.* Journal of Rare Earths, 2024. **42**(5): p. 869-878.

23. Li, Y., et al., *Flexible Ce-doped WO3 nanowire arrays with enriched oxygen vacancies for soft-packaged asymmetric supercapacitor.* Applied Surface Science, 2024. **655**: p. 159616.
24. Dong, L., et al., *Influence of the sintering temperature on the electrical properties of Ce-doped WO3 ceramics prepared from nano-powders.* Journal of Physics D: Applied Physics, 2007. **40**(8): p. 2573.
25. Diao, Q., et al., *Highly sensitive ethanol sensor based on Ce-doped WO3 with raspberry-like architecture.* Materials Research Express, 2020. **7**(11): p. 115012.
26. Gregory, N., *Elements of X-ray diffraction.* Journal of the American Chemical Society, 1957. **79**(7): p. 1773-1774.
27. Naaz, F., et al., *Unraveling the chemoselective catalytic, photocatalytic and electrocatalytic applications of copper supported WO3 nanosheets.* Catalysis Communications, 2023. **178**: p. 106678.
28. Djouadi, D., et al., *Structural and optical characterizations of (Cu, Ce) iso-co-doped ZnO aerogel structures grown in supercritical ethanol.* Journal of Porous Materials, 2019. **26**: p. 755-763.
29. AlKhoori, A.A., et al., *Cu, Sm co-doping effect on the CO oxidation activity of CeO2. A combined experimental and density functional study.* Applied Surface Science, 2020. **521**: p. 146305.
30. Diego-Rucabado, A., et al., *Visible light active Ce-doped and Cu–Ce co-doped TiO2 nanocrystals and optofluidics for clean alcohol production from CO2.* ACS Sustainable Chemistry & Engineering, 2023. **11**(36): p. 13260-13273.
31. Wang, Z., et al., *Low-temperature NO2-sensing properties and morphology-controllable solvothermal synthesis of tungsten oxide nanosheets/nanorods.* Applied Surface Science, 2016. **362**: p. 525-531.
32. Bhargava, R. and S. Khan, *Fabrication of WO3–reduced graphene oxide (WO3–G) nanocomposite for enhanced optical and electrical properties.* Journal of Materials Science: Materials in Electronics, 2020. **31**(11): p. 8370-8384.
33. Chuang, S.-H., H.-C. Lin, and C.-H. Chen, *Oxygen vacancy relationship to photoluminescence and heat treatment methods in hafnium oxide powders.* Journal of alloys and compounds, 2012. **534**: p. 42-46.
34. Zheng, H., et al., *Nanostructured tungsten oxide–properties, synthesis, and applications.* Advanced Functional Materials, 2011. **21**(12): p. 2175-2196.
35. Wang, B., et al., *Structure and photoluminescence of WO3-x aggregates tuned by surfactants.* Micromachines, 2022. **13**(12): p. 2075.
36. Wang, B., et al., *Nanostructure conversion and enhanced photoluminescence of vacancy engineered substoichiometric tungsten oxide nanomaterials.* Materials Chemistry and Physics, 2021. **262**: p. 124311.
37. Kaur, J., et al., *Sensitive and selective acetone sensor based on Gd doped WO3/reduced graphene oxide nanocomposite.* Sensors and Actuators B: Chemical, 2018. **258**: p. 1022-1035.
38. Ning, L., et al., *Electronic properties and 4f→ 5d transitions in Ce-doped Lu 2 SiO 5: a theoretical investigation.* Journal of Materials Chemistry, 2012. **22**(27): p. 13723-13731.
39. Yang, C., et al., *The coupled effect of oxygen vacancies and Pt on the photoelectric response of tungsten trioxide films.* Journal of Materials Chemistry C, 2014. **2**(44): p. 9467-9477.
40. Baishya, K., et al., *Graphene-mediated band gap engineering of WO 3 nanoparticle and a relook at Tauc equation for band gap evaluation.* Applied Physics A, 2018. **124**: p. 1-6.

41. Najafi-Ashtiani, H., et al., *Structural, optical and electrical properties of WO 3–Ag nanocomposites for the electro-optical devices.* Applied Physics A, 2018. **124**: p. 1-9.
42. Haroon, A. and A.S. Ahmed, *An insight into the microstructural properties and dielectric behavior of Ce doped WO3 nanoparticles.* Physica B: Condensed Matter, 2023. **657**: p. 414798.
43. Chansatidkosol, S., et al., *Preparation and assessment of poly (methacrylic acid-coethylene glycol dimethacrylate) as a novel disintegrant.* Tropical Journal of Pharmaceutical Research, 2018. **17**(8): p. 1475-1482.
44. Ramkumar, S. and G. Rajarajan, *A comparative study of humidity sensing and photocatalytic applications of pure and nickel (Ni)-doped WO 3 thin films.* Applied Physics A, 2017. **123**: p. 1-8.
45. Choudhury, S.P. and U.T. Nakate, *Study of improved VOCs sensing properties of boron nitride quantum dots decorated nanostructured 2D-ZnO material.* Ceramics International, 2022. **48**(19): p. 28935-28941.
46. Luo, S., et al., *Gas-sensing properties and complex impedance analysis of Ce-added WO3 nanoparticles to VOC gases.* Solid-state electronics, 2007. **51**(6): p. 913-919.
47. Balasubramani, V., et al., *Recent advances in electrochemical impedance spectroscopy based toxic gas sensors using semiconducting metal oxides.* Journal of the Electrochemical Society, 2020. **167**(3): p. 037572.
48. Hinojo, A., et al., *ZnO sintering aid effect on proton conductivity of BaCe0. 6Zr0. 3Y0. 1O3-δ electrolyte for hydrogen sensors.* Ceramics International, 2024.
49. Sharma, N. and S.P. Choudhury, *Gas sensing using metal oxide semiconductor doped with rare earth elements: A review.* Materials Science and Engineering: B, 2024. **307**: p. 117505.
50. Jia, P., et al., *High-efficient and selective hydrogen gas sensor based on bimetallic Ag/Cu nanoparticles decorated on In2O3: Experimental and DFT calculation.* International Journal of Hydrogen Energy, 2025. **101**: p. 15-25.
51. Liang, J., et al., *Room temperature NO2 sensing properties of Au-decorated vanadium oxide nanowires sensor.* Ceramics International, 2018. **44**(2): p. 2261-2268.
52. Xue, S., et al., *Improving gas-sensing performance based on MOS nanomaterials: a review.* Materials, 2021. **14**(15): p. 4263.